\documentclass[10pt,letterpaper,twocolumn]{article} 

\usepackage{ol2}
\usepackage[draft]{hyperref}
\usepackage{amsmath}
\usepackage{setspace}
\begin{document}

\twocolumn[ 
\title{Fiber-based Radio Frequency Dissemination for Branching Networks with Passive Phase Noise Cancellation}


\author{Y. Bai,$^{1,3}$ B. Wang,$^{1,2,*}$,  C. Gao,$^{1,2}$ J. Miao,$^{1,3}$, X. Zhu,$^{1,3}$, and L. J. Wang$^{1,2,3,4*}$}

\address{
$^1$Joint Institute for Measurement Science, Beijing 100084, China
\\
$^2$Department of Precision Instruments, Tsinghua University, Beijing 100084, China\\
$^3$Department of Physics, Tsinghua University, Beijing 100084, China\\$^4$National Institute of Metrology, Beijing 100013, China\\
$^*$Corresponding author: bo.wang@tsinghua.edu.cn, lwan@tsinghua.edu.cn
}

\begin{abstract}We demonstrate a new fiber-based radio frequency dissemination scheme for branching networks. Without any phase controls on RF signals or usages of active feedback locking components, the highly stable reference frequency signal can be delivered to several remote sites simultaneously and independently. Relative frequency stability of $6\times10^{-15}/s$ and $7\times 10^{-17}/10^4s$ is obtained for 10km dissemination. The proposed low cost and scalable method can be applied to some large-scale frequency synchronization networks.\end{abstract}

 ] 

\begin{figure}[htb]
\centerline{\includegraphics[width=9 cm]{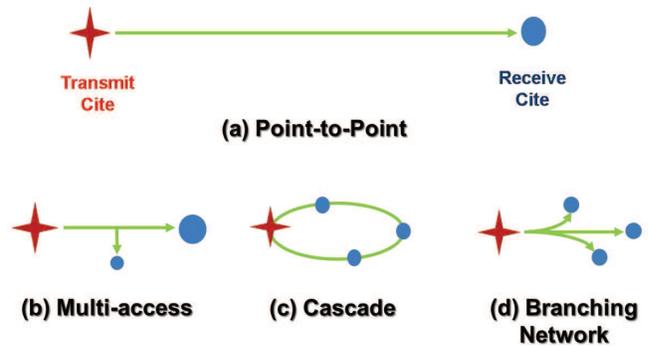}}
\caption{(Color online) Topological structure diagram of different fiber based frequency dissemination schemes.
(a)The conventional point-to-point dissemination scheme.
(b)Muti-access at an arbitrary point along the fiber link.
(c)Cascade dissemination scheme with relay stations.
(d)Branching fiber network dissemination scheme.}
\label{fig1}
\end{figure}

Over the past two decades, highly stable dissemination of time and frequency signals via optical-fiber links has developed considerably and shown broad application prospects\cite{Ma, Levine, Warrington}. Different schemes, such as optical frequency dissemination\cite{Predehl,Lopez}, radio frequency (RF) dissemination\cite{fujieda, Wang}, and optical frequency comb signal dissemination\cite{Marra, Zhao} have been proposed and demonstrated, respectively. However, almost all of these schemes have a common ``point-to-point'' structure, as shown in Figure~\ref{fig1}(a), namely there is only one receive site corresponding to single transmit site. Despite of its high stability, the limited accessibility largely limits further development of the fiber based frequency dissemination technology.

To overcome this mainly drawback, fiber-based multi-access ultrastable radio and optical frequency dissemination schemes have been proposed and demonstrated\cite{Grosche,Gao,Bai,poland,Grosche2,france}. Using this method, highly synchronized RF modulation signal or the optical signal itself can be recovered at any arbitrary points along the fiber link, as shown in Figure~\ref{fig1}(b). To extend the frequency dissemination distance, cascaded frequency disseminations (Figure~\ref{fig1}(c)) have been demonstrated\cite{fujieda2}. 
Along with developments of related technologies and improvements of observation accuracy, more and more large scale scientific and engineering facilities such as radio astronomical observation\cite{radar, VLBI, Cliche} and deep space navigation (DSN) network\cite{NASA, NASA2} require dissemination of reference frequency signals from a certain center site to multiple remote sites over a branching fiber network, as shown in Figure~\ref{fig1}(d). Take the Square Kilometer Array (SKA) project as an example, which represents one of the largest and most challenging timing and synchronization network today\cite{SKA}. Thousands of antennas are quasi-randomly distributed in a broadly symmetric array around the core, at where the reference clock ensemble is located\cite{SKA2}. The key challenge of all fiber based frequency dissemination schemes is how to compensate the fiber induced phase fluctuation. To our knowledge, including the recently demonstrated optical frequency dissemination technique for branching optical-fiber network\cite{schediwy}, almost all existing schemes are developed based on the active phase locking method proposed in 1994 by Ma et al \cite{Ma}. The cost and reliability are main limiting factors for the large-scale application of conventional active fiber-based frequency dissemination schemes.

\begin{figure*}[htb]
\centerline{\includegraphics[width=16cm]{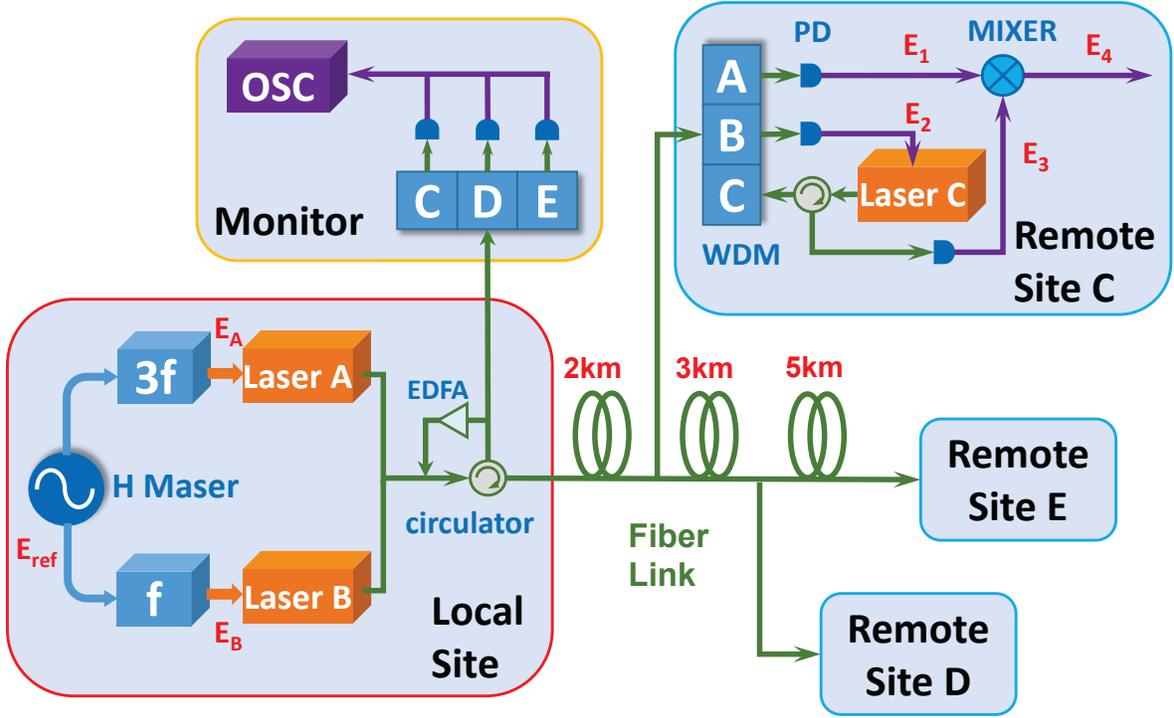}}
\caption{(Color online)Schematic diagram of the fiber based radio frequency dissemination scheme for branching networks with the passive phase noise cancellation method.}
\label{fig2}
\end{figure*}

In this letter we propose and demonstrate a new fiber-based RF dissemination scheme for branching networks. Using the passive phase noise cancelation method, without any phase controls on RF signals or usages of feedback locking components, the highly stable reference frequency signal can be delivered to remote sites simultaneously and independently. For 10 km distance dissemination, relative frequency stability of $6\times10^{-15}/s$ and $7\times 10^{-17}/10^4s$ is obtained. This simple and scalable scheme can be used to construct the frequency synchronization network of DSN and SKA.

Figure~\ref{fig2} shows the schematic diagram of the passive frequency dissemination for branching network experiment. As a laboratorial demonstration, the branching network consists of one local site, three remote sites (C, D, E), and three fiber spools with lengths of 2km, 3km and 5 km. Consequently, remote sites C, D, E are 2km, 5km and 10km away from the local site, respectively. For the convenience of relative frequency stability measurement, the whole system is located at the same lab. Considering the requirements of status monitor and error diagnose in practical applications, we also add an additional monitor part. The optical power and RF waveforms of all remote sites can be monitored at the center station.

At local site, the 100MHz frequency signal of a commercial Hydrogen maser is employed as the reference frequency of whole branching network. It can be expressed as $E_{ref}=V_{ref}cos(\omega_{ref} t+\phi_{ref})$. Two phase-locked dielectric resonant oscillators (PDRO) with frequencies of 3 GHz and 1 GHz are phase locked to $E_{ref}$, and can be expressed as $E_A=V_Acos(3\omega t+3\phi_0)$ and $E_B=V_Bcos(\omega t+\phi_0)$, respectively. The frequency and phase of $E_A$ and $E_B$ have certain triple relationships. They are used to modulate the amplitude of laser A and B, whose wavelengths are 1542nm and 1547nm, respectively. After passing a fiber coupler and a circulator, two modulated laser signals are coupled into the same fiber link and delivered to all remote sites.

The structures of remote sites are almost the same except the wavelengths of laser modules. Here, we choose the remote site C to explain the concept. At remote site C, a wavelength division multiplexer (WDM) is used to separate two received laser signals. The disseminated rf signals can be recovered by two high speed photo-diodes (PD). They can be expressed as

 \begin{equation}
E_1=V_1cos(3\omega t+3\phi_0+3\phi_p),
\end{equation}
 \begin{equation}
E_2=V_2cos(\omega t+\phi_0+\phi_p),
\end{equation}
where $\phi_p$ represents the fiber link induced phase fluctuation for the 1GHz signal $E_B$.  As mentioned above, $E_A$ and $E_B$ have certain triple relationships for both frequency and phase, and they are disseminated in the same fiber simultaneously. Consequently, the fiber link induced phase fluctuation for the 3GHz signal $E_A$ is $3\phi_p$. 
The recovered frequency signal $E_2$ is used to modulate the amplitude of another laser C, whose wavelength is 1550nm. The modulated laser signal is sent back to local site and return to remote site C again with the help of two fiber circulators.  After detected by another PD, the recovered 1 GHz frequency signal can be expressed as

 \begin{equation}
E_3=V_3cos(\omega t+\phi_0+3\phi_p).
\end{equation}

By simply mixing down the signals $E_1$ and $E_3$, we can obtain a 2 GHz signal

 \begin{equation}
E_4=V_4cos(2\omega t+2\phi_0).
\end{equation}
We can see the fiber induced phase fluctuation $\phi_p$ has been passively compensated.

To distinguish the round trip signals of each remote site, the laser modules' wavelengths inside remote sites D and E are chosen as 1552nm and 1555nm, respectively.  Thanks to the mature dense wavelength division multiplexing (DWDM) technology, laser carriers' wavelengths can be as near as 0.4 nm. This will significantly reduce the chromatic dispersion impact on the dissemination stability and expand the quantity of remote sites to dozens or even hundreds.

\begin{figure}[htb]
\centerline{\includegraphics[width=9cm]{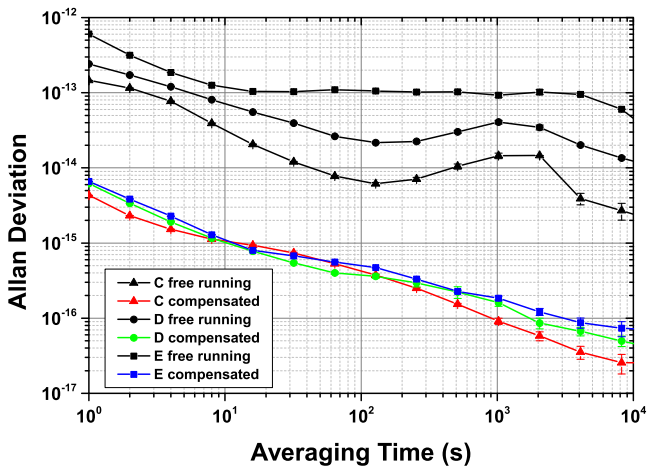}}
\caption{(Color online) Measured relative frequency stabilities of frequency signals at three remote sites C, D and E with and without phase noise passively compensated. }
\label{fig3}
\end{figure}

Through mixing $E_A$ and $E_B$ at local site, we can get $E_{mix}=V_{mix}cos(2\omega t+2\phi_0)$. To measure the dissemination stability of the compensated fiber link, we just measure the relative frequency stability of $E_{mix}$ and $E_4$ and analyze the phase error. We also compare $E_A$ and $E_1$ to measure dissemination stability of the free running fiber link. Similar procedures are taken for remote sites D and E. In order to verify that all remote sites are independent and the proposed scheme is feasible for the application of branching networks, all these measurements are carried out simultaneously.  Figure~\ref{fig3} shows the measured Allan Deviation results. Using the passive phase noise cancelation method, dissemination stability of $6\times10^{-15}/$s and $7\times10^{-17}/10^4s$ is achieved for remote site E (10km away from the local site). While for the free-running fiber link, it has dissemination stability of $6\times10^{-13}/$s and $6\times10^{-14}/10^4s$, respectively. We also notice that, for the free running fiber links, dissemination stabilities decrease along with the increase of transmission distance. While for the passively compensated cases, dissemination stabilities of the three remote sites can achieve the same order of magnitude.

\begin{figure}[htb]
\centerline{\includegraphics[width=8.5 cm]{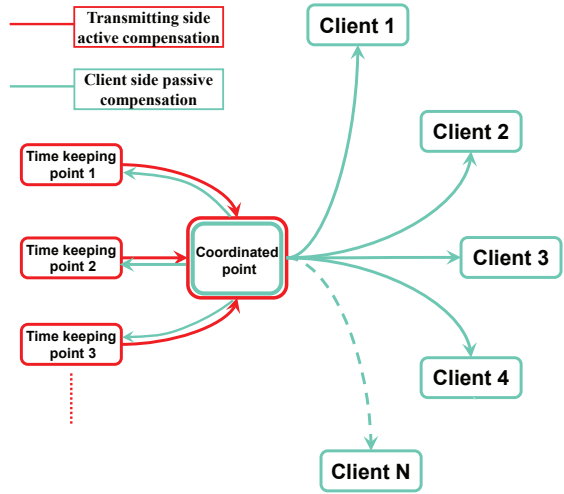}}
\caption{(Color online) Schematic diagram of a regional time and frequency synchronization network.}
\label{fig4}
\end{figure}

As a primary application example, we take a brief introduction of the Beijing regional time and frequency synchronization network which is under construction. As shown in Fig.~\ref{fig4}, the synchronization network can be separated into three parts - time keeping stations, coordinating station and client sites. The time keeping points are the institutes which maintain the national or regional official times. Using the conventional active compensation scheme\cite{Wang}, a time keeping station can perform frequency comparison, and transmit its time-frequency signal to a central coordinating station. Here, the received time-keeping signals can be compared in real time to generate a coordinated time-frequency signal. Then the coordinated frequency can be disseminated back to each time keeping stations and forward to all user clients via the proposed passive compensation method. As one transmitting module can be linked with multiple remote sites, the expansion of future dissemination channels (expansion of the clients) will not have significant impact on the coordinating station's basic structure.

In summary, we have demonstrated a new fiber-based, radio frequency dissemination scheme for branching network. Using the method, stable, distributed RF signals can be simultaneously and independently disseminated to many different remote sites from one central station. Some scientific construction projects are expected to benefit from this method.

We acknowledge funding supports from the National Key Scientific Instrument and Equipment Development Projects (No.2013YQ09094303).

\pagebreak
\section*{Informational Fourth Page}
 {\bf Full versions of citations}

\end{document}